\documentclass[twocolumn,english,american]{revtex4-2}
\usepackage[T1]{fontenc}
\usepackage[utf8]{luainputenc}
\usepackage{amsmath}
\usepackage{babel}
\begin{document}
\title{The Wilson-Fisher Fixed point revisited: importance of the form of
the cutoff}
\author{Garry Goldstein$^{1}$}
\address{$^{1}$garrygoldsteinwinnipeg@gmail.com}
\begin{abstract}
In this work we re-examine the Wilson Fisher fixed point. We study
Wilsonian momentum space renormalization group (RG) flow for various
forms of the cutoff. We show that already at order $\left(4-d\right)^{1}$,
where $d$ is the dimension of the $\phi^{4}$ theory, there are changes
to the position of the fixed point and the direction of irrelevant
coupling parameters. We also show in a multi-flavor $\phi^{4}$ model
that symmetries of the Lagrange function can be destroyed if the different
flavors have different cutoffs (that is the Lagrangian can flow to
a non-symmetric fixed point). Some related comments are made about
a similar situation in parquet RG (pRG). In future works \citep{Goldstein_2024}
we will study Wilsonian RG to order $\left(4-d\right)^{2}$ and find
non-universal critical exponents that depend on the cutoff.
\end{abstract}
\maketitle

\section{Introduction}\label{sec:Introduction}

One of the most successful approaches to the study of both quantum
and classical phase transitions is the momentum space renormalization
group (RG) \citep{Sachdev_2011,Sachdev_2023,Wilson_1971,Wilson_1971(2),Wilson_1972,Continentino_2017,Kopietz_2010,Parisi_1988,Ramond_2001,Wilson_1974,Zinn_Justin_2002,Zinn_Justin_2007}.
The momentum space renormalization group provides the mathematical
underpinnings of both scaling and critical phenomena. These ideas
were developed by Kadanoff \citep{Kadanoff_1966}, Wilson \citep{Wilson_1971,Wilson_1971(2),Wilson_1972}
and others \citep{Pathria_2011,Continentino_2017,Sachdev_2011,Sachdev_2023}
into powerful mathematical tools for tackling the physics problems
associated with scaling, criticality and phase transitions. The main
idea supporting momentum space RG is due to Kadanoff \citep{Kadanoff_1966}
is that close to a phase transition a critical system becomes scale
invariant. That is, there are very few relevant parameters (say the
mass in $\phi^{4}$ theory) which control the distance between the
theory (system) and the scale invariant critical theory where all
correlation lengths are infinite. Under a rescaling (integrating out)
of various degrees of freedom these relevant parameters rescale in
a specific way and the system looks like a scaled system with renormalized
parameters. In the simplest case when there is only one relevant parameter,
under a rescaling by a dilation factor of $l,$ the relevant parameter
$t$ and the correlation length $\xi$ transform as \citep{Pathria_2011,Continentino_2017,Sachdev_2011}:
\begin{equation}
t\rightarrow t\cdot l^{\nu},\:\xi\rightarrow\xi\cdot l\label{eq:Scaling}
\end{equation}
For some exponent $\nu$, which Wilsonian RG can compute. In this
case we see that we may read off the correlation length from the value
of the relevant parameter $t$:
\begin{equation}
tl^{\nu}\sim1\rightarrow\:\xi\sim1,\Rightarrow\:\xi\sim t^{-1/\nu}\label{eq:Scaling_critical}
\end{equation}
The idea due to Wilson \citep{Wilson_1971,Wilson_1971(2),Wilson_1972}
and others was to systematically, in momentum space, integrate out
the short wavelength degrees of freedom thereby mimicking a rescaling
transformation and follow mathematically how relevant and also irrelevant
parameters transform. As such Wilson \citep{Wilson_1971,Wilson_1971(2),Wilson_1972,Sachdev_2011}
and others were able to compute $\nu$ and other critical parameters
through perturbative momentum space RG calculation. 

There are roughly three different types of momentum shell RG: 1) Wilsonian
RG \citep{Wilson_1971,Wilson_1971(2),Wilson_1972,Wilson_1974,Sachdev_2011,Pathria_2011,Continentino_2017},
2) field theoretic RG \citep{Parisi_1988,Ramond_2001,Zinn_Justin_2002,Zinn_Justin_2007},
3) functional RG \citep{Kopietz_2010}. In Wilsonian RG there is a
cutoff, it is rescaled and the effect of integrating out the high
degrees of freedom in the effective Lagrange function is accounted
for systematically up to some loop order \citep{Pathria_2011,Wilson_1971,Wilson_1971(2),Wilson_1972,Wilson_1974,Sachdev_2011,Continentino_2017}.
In field theoretic RG the theory is regularized, often by dimensional
regularization - where divergent integrals are analytically continued
from dimensions where they converge, renormalization conditions are
demanded - where specific values are set to Green's functions at certain
momenta - and the Callan-Symanzik equation is set up to compute Green's
functions at other momenta \citep{Parisi_1988,Ramond_2001,Zinn_Justin_2002,Zinn_Justin_2007}.
In FRG the propagators are infinitely massive below a certain cutoff
which is then rescaled and the effective action is computed as a function
of the cutoff \citep{Kopietz_2010}. In each case a regularization,
cutoff is needed to make the theory finite and control the divergences
of various Feynman diagrams. The effect of the nature of this cutoff
on the RG procedure has been studied with mixed success. For the Kondo
model at two loop order the Bethe Ansatz solution and the numerical
Wilsonian RG solution seem to differ, perhaps because of the nature
of the cutoff \citep{Andrei_1983,Coleman_2016,Hewson_1993}. For field
theoretic RG different regulators have been compared with favorable
results indicating similar physical properties for all cutoffs \citep{Kopietz_2010,Zinn_Justin_2002}.
For FRG there have been numerous studies on the subject matter \citep{Liao_2000,Litim_2000,Litim_2001,Gaite_2023}
with some universality observed \citep{Liao_2000,Litim_2000,Litim_2001}
though higher loop corrections seem to dependent on the form of the
cutoff \citep{Gaite_2023}. 

Wilsonian RG is arguably the simplest of the three forms of RG. It
has been used to study $\varphi^{4}$ theory, $O\left(n\right)$ models,
non-linear $\sigma$ models, numerically the Kondo problem, sine gordon
theory, Hertz-Millis theory \citep{Sachdev_2011,Pathria_2011,Wen_2004,Giamarchi_2003,Fradkin_2013,Continentino_2017,Hewson_1993}.
Here we study the effect of the nature of the cutoff on Wilsonian
RG within the context of the $\varphi^{4}$ theory in the form of
a $4-d$ expansion. Here we claim Wilson and others were somewhat
careless as to the nature of what constitutes high frequency and high
energy modes as they introduced a hard cutoff to separate high frequency
and low frequency modes. Here following the paradigmatic example of
the Wilson Fisher fixed point \citep{Wilson_1972}, which works well
for $\phi^{4}$ theory in the limit that we are working at nearly
four dimensions $4-d\ll1$, we show that the choice of hard cutoff
is a poor unjustified choice by comparing it top other choices. By
varying the form of the cutoff (see Eq. (\ref{eq:Soft_cutoff_general}))
we explicitly show that we can obtain different values for the fixed
point couplings (see Eq. (\ref{eq:Fixed_general})) and modify the
various relevant and irrelevant directions as well as how they scale
(see Eq. (\ref{eq:Irrelevant_direction})). This invalidates, in part,
Wilson's ideas as how to practically compute within the momentum space
RG flow the scaling transformations introduced by Kadanoff as it is
unclear which cutoff is best. This has far ranging implications for
modern day condensed matter \citep{Continentino_2017,Sachdev_2011,Sachdev_2023}.
Condensed matter systems often comes with a practical cutoff associated
with the lattice regularization (no modes with wave vector higher
then the inverse lattice spacing are needed for most calculations).
However this regularization is often considered far too difficult
for practical calculations and a simpler ``spherical cow'' regularization
is chosen whereupon momentum space RG is performed within this regularization
\citep{Continentino_2017,Sachdev_2011,Sachdev_2023}. The results
of this work, where we choose a model ($\phi^{4}$ theory) where it
is practically possible to compare several different regularizations
and obtain different results already at one loop level, casts strong
doubts as to the validity of the procedure commonly carried out by
many condensed matter physicists. Furthermore we show that variations
in the cutoff for different field flavors, in multi-flavor $\phi^{4}$
models, can lead to changes of the symmetry of the system whereby
a symmetric (we chose the example of $O\left(2\right)$ symmetric)
action can flow to non symmetric fixed points (see Eq. (\ref{eq:RG_flow_symmetries})).
As such the symmetries and minimal Lagrange function relevant to the
order parameter action can change due to different cutoffs \citep{Continentino_2017,Sachdev_2023}.
We also point out that a similar effect happens in parquet RG (pRG)
and is arguably an even worse concern. In future works we will go
to order $\left(4-d\right)^{2}$ to further confirm these results
\citep{Goldstein_2024}.

\section{Main Idea}\label{sec:Main-Idea}

We consider the $\phi^{4}$ theory with the action being given by:
\begin{equation}
S=-\int d^{d}x\left[\phi\left(x\right)\frac{1}{2}\left[-\nabla^{2}+r\right]\phi\left(x\right)+\frac{1}{4!}g\phi^{4}\left(x\right)\right]\label{eq:Action_dimensionful}
\end{equation}
We now regularize the theory by replacing: 
\begin{equation}
\int d^{d}\mathbf{x}\rightarrow\int\frac{d^{d}\mathbf{k}}{\left(2\pi\right)^{d}}F^{1/2}\left(\frac{\left|\mathbf{k}\right|}{\Delta}\right)\label{eq:Soft_cutoff_general}
\end{equation}
For some large cutoff $\Delta$ with $F\left(0\right)=1$ and $F\left(\infty\right)\rightarrow0$
rapidly and the square root is for future convenience. It is straightforward
to see that the main effect of this regulator on Feynman diagrams
is to change the propagator: 
\begin{equation}
\frac{1}{\mathbf{Q}^{2}+r}\rightarrow\frac{F\left(\frac{\left|\mathbf{Q}\right|}{\Delta}\right)}{\mathbf{Q}^{2}+r}\label{eq:Transform}
\end{equation}
Here $\mathbf{Q}$ is the momentum. This makes the procedure similar
to FRG \citet{Kopietz_2010}. The main interest in this work is how
the form of the cutoff effects the Wilson Fisher fixed point \citep{Wilson_1972}
to order $\left(4-d\right)^{1}$, for order $\left(4-d\right)^{2}$
see \citet{Goldstein_2024}. Now we see that the action is dimensionless
so we may write that the scaling dimensions are given by $\left[\phi\right]=\frac{d-2}{2}$,
and $\left[g\right]=4-d$. As such we can rewrite Eq. (\ref{eq:Action_dimensionful})
as: 
\begin{equation}
S=\int d^{d}x\left[\phi^{\dagger}\left(x\right)\left[-\nabla^{2}+\Delta^{2}R\right]\phi\left(x\right)+\frac{1}{4!}\Delta^{4-d}G\phi^{4}\left(x\right)\right]\label{eq:Action_dimensionless}
\end{equation}
with $R$ and $G$ dimensionless. Now we perform a step of RG whereby:
\begin{equation}
\Delta\rightarrow\Delta\left(1-\varepsilon\right)\label{eq:Rescaling}
\end{equation}
To compute the new action we first we do the tree level RG step (that
is to order $\left(4-d\right)^{0}$) where:
\begin{align}
\Delta^{2}R & =\Delta^{2}\left(1-\varepsilon\right)^{2}R'\nonumber \\
\Rightarrow R^{'}= & R\left(1+2\varepsilon\right)\label{eq:Tree_level_I}
\end{align}
\begin{align}
\Delta^{4-d}G & =\Delta^{4-d}\left(1-\varepsilon\right)^{4-d}G'\nonumber \\
\Rightarrow G^{'}= & G\left(1+\varepsilon\left(4-d\right)\right)\label{eq:Tree_level_II}
\end{align}
 Now we perform RG to one loop level where we have that:
\begin{widetext}
\begin{align}
\Delta^{2}R & \rightarrow\Delta^{2}R-\varepsilon\Delta^{4-d}G\frac{V\left(S^{d-1}\right)}{2\left(2\pi\right)^{d}}\int_{0}^{\infty}dk\frac{\partial}{\partial\varepsilon}\left[F\left(\frac{k}{\Delta}\left(1+\varepsilon\right)\right)\right]\frac{1}{k^{2}+\Delta^{2}R}k^{d-1}\nonumber \\
\frac{1}{4!}\Delta^{4-d}G & \rightarrow\frac{1}{4!}\Delta^{4-d}G+3\varepsilon\left(\Delta^{4-d}G\right)^{2}\frac{V\left(S^{d-1}\right)}{2\left(2\pi\right)^{d}}\int_{0}^{\infty}dk\frac{\partial}{\partial\varepsilon}\left[F^{2}\left(\frac{k}{\Delta}\left(1+\varepsilon\right)\right)\right]\left[\frac{1}{k^{2}+\Delta^{2}R}\right]^{2}k^{d-1}\label{eq:One_loop_RG_step}
\end{align}
\end{widetext}

As such we have that: 
\begin{align}
 & -\int_{0}^{\infty}dk\frac{\partial}{\partial\varepsilon}\left[F\left(\frac{k}{\Delta}\left(1+\varepsilon\right)\right)\right]\frac{1}{k^{2}+\Delta^{2}R}k^{d-1}\nonumber \\
 & =\int_{0}^{\infty}dk\left(\frac{k}{\Delta}\right)\frac{1}{k^{2}+\Delta^{2}R}k^{d-1}\frac{\partial}{\partial\left(\frac{k}{\Delta}\right)}F\left(\frac{k}{\Delta}\right)\label{eq:Simplifications_I}
\end{align}
\begin{align}
 & \int_{0}^{\infty}dk\frac{\partial}{\partial\varepsilon}\left[F^{2}\left(\frac{k}{\Delta}\left(1+\varepsilon\right)\right)\right]\left[\frac{1}{k^{2}+\Delta^{2}R}\right]^{2}k^{d-1}\nonumber \\
 & =-\int_{0}^{\infty}dk\frac{k}{\Delta}\left[\frac{1}{k^{2}+\Delta^{2}R}\right]^{2}k^{d-1}\frac{\partial}{\partial\left(\frac{k}{\Delta}\right)}F\left(\frac{k}{\Delta}\right)\label{eq:Simplifications_II}
\end{align}
We now introduce $K=\frac{k}{\Delta}$ and obtain:
\begin{align}
 & \int_{0}^{\infty}dk\left(\frac{k}{\Delta}\right)\frac{1}{k^{2}+\Delta^{2}R}k^{d-1}\frac{\partial}{\partial\left(\frac{k}{\Delta}\right)}F\left(\frac{k}{\Delta}\right)\nonumber \\
 & =\Delta^{d-2}\int_{0}^{\infty}dK\frac{\partial}{\partial K}F\left(K\right)\frac{1}{K^{2}+R}K^{d}\label{eq:Change_variable_I}
\end{align}
\begin{align}
 & \int_{0}^{\infty}dk\frac{k}{\Delta}\left[\frac{1}{k^{2}+\Delta^{2}R}\right]^{2}k^{d-1}\frac{\partial}{\partial\left(\frac{k}{\Delta}\right)}F^{2}\left(\frac{k}{\Delta}\right)\nonumber \\
 & =\Delta^{d-4}\int_{0}^{\infty}dK\frac{\partial}{\partial K}F^{2}\left(K\right)\left[\frac{1}{K^{2}+R}\right]^{2}K^{d}\label{eq:Change_variable_II}
\end{align}
 Now we have that: 
\begin{align}
R & \rightarrow R-\varepsilon G\frac{V\left(S^{d-1}\right)}{\left(2\pi\right)^{d}}\int_{0}^{\infty}dK\frac{\partial}{\partial K}F\left(K\right)\frac{1}{K^{2}+R}K^{d}\nonumber \\
 & \cong R+\varepsilon G\frac{V\left(S^{d-1}\right)}{\left(2\pi\right)^{d}}I_{1}^{F,d}\left(R\right)\nonumber \\
G & \rightarrow G+3\varepsilon G^{2}\frac{V\left(S^{d-1}\right)}{\left(2\pi\right)^{d}}\int_{0}^{\infty}dK\frac{\partial}{\partial K}F^{2}\left(K\right)\left[\frac{1}{K^{2}+R}\right]^{2}K^{d}\nonumber \\
 & \cong G-3\varepsilon G^{2}\frac{V\left(S^{d-1}\right)}{\left(2\pi\right)^{d}}I_{2}^{F,d}\left(R\right)\label{eq:Transform-1}
\end{align}
As such combining with Eq. (\ref{eq:Tree_level_RG}) we obtain:

\begin{align}
\frac{\partial R}{\partial\varepsilon} & =2R+\frac{1}{2}G\frac{V\left(S^{3}\right)}{\left(2\pi\right)^{4}}I_{1}^{F,d}\left(R\right)\nonumber \\
\frac{\partial G}{\partial\varepsilon} & =\left(4-d\right)G-\frac{3}{2}G^{2}\frac{V\left(S^{3}\right)}{\left(2\pi\right)^{4}}I_{2}^{F,d}\left(R\right)\label{eq:RG_flow_final}
\end{align}
Where 
\begin{align}
I_{1}^{F,d}\left(R\right) & =-\int_{0}^{\infty}dK\frac{\partial}{\partial K}F\left(K\right)\frac{1}{K^{2}+R}K^{d}\nonumber \\
I_{2}^{F,d}\left(R\right) & =-\int_{0}^{\infty}dK\frac{\partial}{\partial K}F^{2}\left(K\right)\left[\frac{1}{K^{2}+R}\right]^{2}K^{d}\label{eq:Integrals}
\end{align}
Working only to order $4-d\ll1$ and working only to order $\left(4-d\right)$
(that is ignoring order $\left(4-d\right)^{2}$ terms) we need only
consider the integrals:
\begin{align}
I_{1}^{F,4}\left(0\right)=-\int_{0}^{\infty}dK\frac{\partial}{\partial K}F\left(K\right)K^{2} & =2\int_{0}^{\infty}dKF\left(K\right)K\nonumber \\
I_{2}^{F,4}\left(0\right)=-\int_{0}^{\infty}dK\frac{\partial}{\partial K}F^{2}\left(K\right) & =1\label{eq:Integrals-1}
\end{align}
See supplementary online information \citep{Supplement_2024}. We
now look for a fixed point and again work only to order $4-d\ll1$
then:
\begin{equation}
G^{fix}=\frac{2\left(4-d\right)\left(2\pi\right)^{4}}{3V\left(S^{3}\right)},\:R^{fix}=-\frac{\left(4-d\right)I_{1}^{F,4}\left(0\right)}{6}\label{eq:Fixed_general}
\end{equation}
We note that this is not the value of the Willson Fisher fixed point
given by \citep{Sachdev_2011,Continentino_2017,Wilson_1972}: 
\begin{equation}
G_{WF}^{fix}=\frac{2\left(4-d\right)\left(2\pi\right)^{4}}{3V\left(S^{3}\right)},\:R_{WF}^{fix}=-\frac{\left(4-d\right)}{6}\label{eq:WF_Fixed}
\end{equation}
We see that the exact position of the fixed point has moved base on
the value of $I_{1}^{F,4}\left(0\right)$ \citep{Supplement_2024}
as such the critical theory will change \citep{Sachdev_2011}.

\section{Critical exponents}\label{sec:Critical-exponents}

Let us recall Eq. (\ref{eq:RG_flow_final}). Now we write 
\begin{align}
G & =G_{fix}+\delta G\nonumber \\
R & =R_{fix}+\delta R\label{eq:Shift}
\end{align}
From which we see that:
\begin{equation}
\frac{\partial}{\partial\varepsilon}\left(\begin{array}{c}
\delta R\\
\delta G
\end{array}\right)\cong\left(\begin{array}{cc}
2+\frac{2\left(4-d\right)}{3} & \frac{1}{2}\frac{V\left(S^{3}\right)}{\left(2\pi\right)^{4}}I_{1}^{F,4}\left(0\right)\\
0 & -\left(4-d\right)
\end{array}\right)\left(\begin{array}{c}
\delta R\\
\delta G
\end{array}\right)\label{eq:Critical}
\end{equation}
Because the eigenvalues of the matrix in Eq. (\ref{eq:Critical})
do not depend on $F$ we recover the Wilson Fisher critical exponents
to order $4-d$, however because of the changes to $G_{fix}$, $R_{fix}$
there will be changes to order $\left(4-d\right)^{2}$ \citep{Goldstein_2024}.
However the irrelevant eigenvector changes to:
\begin{equation}
\left(\begin{array}{cc}
2+\frac{2\left(4-d\right)}{3} & \frac{1}{2}\frac{V\left(S^{3}\right)}{\left(2\pi\right)^{4}}I_{1}^{F,4}\left(0\right)\\
0 & -\left(4-d\right)
\end{array}\right)\left(\begin{array}{c}
a\\
b
\end{array}\right)=-\left(4-d\right)\left(\begin{array}{c}
a\\
b
\end{array}\right)\label{eq:eigenvalue_equation}
\end{equation}
whereby: 
\begin{equation}
\frac{a}{b}=-\frac{\frac{1}{2}\frac{V\left(S^{3}\right)}{\left(2\pi\right)^{4}}I_{1}^{F,4}\left(0\right)}{2+\frac{5}{3}\left(4-d\right)}\label{eq:Irrelevant_direction}
\end{equation}
so there is an explicit change to the irrelevant vector depending
on the values of $\frac{\frac{1}{2}\frac{V\left(S^{3}\right)}{\left(2\pi\right)^{4}}I_{1}^{F,4}\left(0\right)}{2+\frac{5}{3}\left(4-d\right)}.$
As such the critical flow changes due to various types of cutoffs
already at leading order and not just the position of the fixed point.

\section{Emergent symmetries}\label{sec:Symmetry-analysis}

We would like to show that emergent symmetries can also be affected
by the softness of the cutoff not just the value of the fixed point.
We consider the two species $\phi^{4}$ theory given by: 
\begin{align}
S= & -\frac{1}{2}\int d^{d}x\phi_{1}\left(x\right)\left[-\nabla^{2}+r_{1}\right]\phi_{1}\left(x\right)\nonumber \\
 & -\frac{1}{2}\int d^{d}x\phi_{2}\left(x\right)\left[-\nabla^{2}+r_{2}\right]\phi_{2}\left(x\right)\nonumber \\
 & -\frac{1}{4!}\int d^{d}x\sum_{ab}g_{ab}\phi_{a}^{2}\left(x\right)\phi_{b}^{2}\left(x\right)\label{eq:Action_symmetry}
\end{align}
where $a,b=1,2$ with $g_{12}=g_{21}$. Now we regularize the theory
in a different way for each of the fields:
\begin{align}
a=1 & :\int d^{d}x\rightarrow\int\frac{d^{d}k}{\left(2\pi\right)^{d}}F_{1}^{1/2}\left(\frac{k}{\Delta}\right)\nonumber \\
a=2 & :\int d^{d}x\rightarrow\int\frac{d^{d}k}{\left(2\pi\right)^{d}}F_{2}^{1/2}\left(\frac{k}{\Delta}\right)\label{eq:Soft_cutoff_different}
\end{align}
 Now we write a dimensionless variable action: 
\begin{align}
S= & -\frac{1}{2}\int d^{d}x\phi_{1}\left(x\right)\left[-\nabla^{2}+\Delta^{2}R_{1}\right]\phi_{1}\left(x\right)\nonumber \\
 & -\frac{1}{2}\int d^{d}x\phi_{2}\left(x\right)\left[-\nabla^{2}+\Delta^{2}R_{2}\right]\phi_{2}\left(x\right)\nonumber \\
 & -\frac{1}{4!}\int d^{d}x\Delta^{4-d}G_{ab}\phi_{a}^{2}\left(x\right)\phi_{b}^{2}\left(x\right)\label{eq:Action_symmetry_dimensionless}
\end{align}
Now we perform a step of RG as in Eq. (\ref{eq:Rescaling}). We now
find the RG equations of motion to one loop order: keeping only terms
of order $4-d$ (similarly to Eq. (\ref{eq:RG_flow_final}): 
\begin{align}
\frac{\partial}{\partial\varepsilon}R_{1} & =2R_{1}+\frac{1}{2}\frac{V\left(S^{3}\right)}{\left(2\pi\right)^{4}}\left[G_{11}I_{1}^{F_{1},4}\left(0\right)+\frac{1}{3}G_{12}I_{1}^{F_{2},4}\left(0\right)\right]\nonumber \\
\frac{\partial}{\partial\varepsilon}R_{2} & =2R_{2}+\frac{1}{2}\frac{V\left(S^{3}\right)}{\left(2\pi\right)^{4}}\left[G_{22}I_{1}^{F_{2},4}\left(0\right)+\frac{1}{3}G_{12}I_{1}^{F_{1},4}\left(0\right)\right]\nonumber \\
\frac{\partial}{\partial\varepsilon}G_{11} & =\left(4-d\right)G_{11}-\frac{V\left(S^{3}\right)}{\left(2\pi\right)^{4}}\left[\frac{3}{2}G_{11}^{2}+\frac{1}{6}G_{12}^{2}\right]\nonumber \\
\frac{\partial}{\partial\varepsilon}G_{22} & =\left(4-d\right)G_{22}-\frac{V\left(S^{3}\right)}{\left(2\pi\right)^{4}}\left[\frac{3}{2}G_{22}^{2}+\frac{1}{6}G_{12}^{2}\right]\nonumber \\
\frac{\partial}{\partial\varepsilon}G_{12} & =\left(4-d\right)G_{12}\nonumber \\
 & -\frac{V\left(S^{3}\right)}{\left(2\pi\right)^{4}}\left[\frac{3}{4}G_{22}G_{12}+\frac{3}{4}G_{11}G_{12}+\frac{1}{6}G_{12}^{2}\right]\label{eq:RG_flow_symmetries}
\end{align}
Now we see that the fixed point corresponds to 
\begin{align}
G_{11}^{fix} & =G_{12}^{fix}=G_{22}^{fix}=\frac{3\left(4-d\right)}{5\frac{V\left(S^{3}\right)}{\left(2\pi\right)^{4}}}\nonumber \\
R_{1}^{fix} & =-\frac{3\left(4-d\right)}{20}\left[I_{1}^{F_{1},4}\left(0\right)+\frac{1}{3}I_{1}^{F_{2},d}\left(0\right)\right]\nonumber \\
R_{2}^{fix} & =-\frac{3\left(4-d\right)}{20}\left[I_{1}^{F_{2},4}\left(0\right)+\frac{1}{3}I_{1}^{F_{1},d}\left(0\right)\right]\label{eq:Fixed_points}
\end{align}
We see that the $O\left(2\right)$ symmetric point is no longer a
fixed point of the low energy theory. Furthermore, we see that we
can start with a $O\left(2\right)$ symmetric action and under RG
flow to an non-symmetric fixed point. 

\section{Comment on pRG}\label{sec:Comment-on-pRG}

The situation with pRG is worse then with the Wilson Fisher fixed
point with respect to cutoff dependence. We recall that in pRG one
introduces various couplings between different patches \citep{Maiti_2013}
and computes to a certain loop level the changes of these couplings
when integrating out high energy degrees of freedom. The one loop
equations of motion for the various couplings $g_{i}$ are given by
\citep{Maiti_2013}: 
\begin{equation}
\frac{\partial g_{i}}{\partial\Delta}=\sum_{jk}A_{jk}^{i}\left(\Delta\right)g_{j}g_{k}\label{eq:pRG}
\end{equation}
Here $A_{jk}\left(\Delta\right)$ are parameters that depend on the
density of states associated with the various polarization bubbles
\citep{Maiti_2013} and $\Delta$ is an overall flow parameter which
controls how these density of states for the relevant patches is integrated
out. Now we look for blow up solutions \citep{Maiti_2013}
\begin{equation}
g_{i}=\frac{c_{i}}{\Delta-\Delta_{0}}+...\label{eq:Blow_up}
\end{equation}
Now we note then that \citep{Maiti_2013} we have that: 
\begin{equation}
c_{i}=\sum_{jk}A_{jk}^{i}\left(\Delta_{0}\right)c_{j}c_{k}\label{eq:c_i}
\end{equation}
Now we may perform pRG with both smooth and sharp cutoffs and obtain
very different $\Delta_{0}$ and $A_{jk}^{i}\left(\Delta_{0}\right)$
much like in the main part of the paper \citep{Goldstein_2024}. As
such the various $c_{i}$ which control how the various couplings
blow up are modified due to different cutoff dependent $A_{jk}^{i}\left(\Delta_{0}\right)$.
Therefore the various instabilities which depend on the signs and
magnitudes of the $c_{i}$ (see Eq. (\ref{eq:Blow_up}) now depend
on the shape of the cutoff . As such pRG has reliability problems
as different cutoffs can predict different instabilities \citep{Goldstein_2024}.

\section{Conclusions}\label{sec:Conclusions}

In this work we have re-examined the Wilson Fisher fixed point \citep{Wilson_1972}
in the context of more general cutoffs then the hard cutoff considered
in the original Wilson Fisher work \citep{Wilson_1972}. We find that
both the position of the fixed point the direction of irrelevant perturbations
depends explicitly on the form of the cutoff already at order $\left(4-d\right)$
or one loop level and explicit checks will show that the critical
exponents depend on the form of the cutoff at the two loop order $\left(4-d\right)^{2}$
level \citet{Goldstein_2024}. We have shown that emergent symmetries
where the theory flows to actions with high level of symmetry such
as $O\left(2\right)$ depend on the form of the cutoff and if different
flavors of fields have different cutoffs the high symmetry fixed point
can be destroyed so a $O\left(2\right)$ symmetric theory may flow
to a non-symmetric fixed point already at order $4-d$. Since the
nature of the order parameter and other properties of the system depend
on the symmetries of the low energy effective action \citep{Sachdev_2011,Sachdev_2023,Pathria_2011,Continentino_2017}
this calculation shows that the form of the cutoff can influence the
form of the order parameter for the effective theory. We have shown
that the situation with pRG is even more difficult to control then
with the Wilson Fisher fixed point. Indeed for pRG divergences show
up at finite times in the flow parameter \citep{Maiti_2013} where
depending on the nature of the cutoff and how it flows the density
of states $A_{jk}^{i}\left(\Delta_{0}\right)$ can significantly change
whereby which couplings diverge how fast (see Eqs. (\ref{eq:Blow_up})
and (\ref{eq:c_i})) can change which leads to different instabilities
being dominant \citep{Maiti_2013}. In the eyes of the author this
makes pRG highly unreliable as a method to predict which instabilities
dominate depending on which density of states and which initial couplings
are present. In future works it would be of great interest to see
what can be salvaged from momentum space RG type ideas in a highly
controlled reliable way and to study the Wilson Fisher fixed point
for arbitrary cutoffs to order $\left(4-d\right)^{2}$. 

\textbf{Acknowledgements:} The author would like to thank Jose Gaite
for useful discussions.

\appendix

\part*{Supplementary online information}

\section*{Some explicit examples}\label{subsec:Some-explicit-examples}

We will now explicitly show there are no relations for the function
$I_{1}^{F,4}\left(0\right)$. Now consider the functions cutoffs:
\begin{align}
F_{1}\left(K\right) & =\Theta\left(K-1\right)\nonumber \\
F_{2}\left(K\right) & =\exp\left(-K\right)\nonumber \\
F_{3}\left(K\right) & =\exp\left(-K^{2}\right)\nonumber \\
F_{4}\left(K\right) & =\exp\left(-K^{4}\right)\label{eq:Functions}
\end{align}
Here $\Theta$ is the heavy-side function. Then we have that: 
\begin{align}
I_{1}^{F_{1},4}\left(0\right) & =1\nonumber \\
I_{1}^{F_{2},4}\left(0\right) & =2\nonumber \\
I_{1}^{F_{3},4}\left(0\right) & =1\nonumber \\
I_{1}^{F_{4},4}\left(0\right) & =\frac{\sqrt{\pi}}{2}\label{eq:Some_numbers}
\end{align}
As such $I^{F,4}\left(0\right)>0$ is an arbitrary non-universal constant
effecting the fixed point and flow of RG.

\end{document}